\documentstyle[12 pt]{article}

\title{\bf The
$ n $- jet inclusive cross-section in the Hard  Pomeron model}
\author{Mikhail Braun\thanks{
 Permanent address: Dep. High-Energy Physics, University of St. Petersburg,
198904 St.Petersburg, Russia}
\hspace{2mm}and Daniele Treleani \\
Dipartimento di Fisica Teorica dell'Universit\`a and INFN.\\
Trieste, I 34014 Italy}
\def\beq{\begin{equation}}
\def\eeq{\end{equation}}
\def\bea{\begin{eqnarray}}
\def\eea{\end{eqnarray}}
\def\noi{\noindent}

\begin{document}
\maketitle
\medskip
\centerline{\bf Abstract.}
The inclusive production of several jets is studied in the framework
of the hard pomeron model. The average jet momenta are found to be
strongly ordered and growing towards the central c.m. rapidity, as expected.
Strong positive correlations are found for pairs of jets neighbouring
in rapidity, both in the forward-backward and forward-forward
directions.

\newpage
%%%%%%%%%%%%%%%%%%%%%%%%%%%%%%%%%%%%%%%%%%%%%%%%%%%%%%%%%%%%%%%%%%%%%%%%%%%%
\section{ Introduction.}
%%%%%%%%%%%%%%%%%%%%%%%%%%%%%%%%%%%%%%%%%%%%%%%%%%%%%%%%%%%%%%%%%%%%%%%%%%%%
Recent experimental data on low
$ x $ behaviour of the proton structure functions at HERA [ 1 ] seem to support
the idea that this behaviour is governed by an exchange of a
hard ("BFKL") pomeron. However a completely inclusive character of the
structure function measurements makes it very difficult to draw
definite conclusions on this point. More information is likely to be
obtained from less inclusive experiments in  which the structure of the
produced particles spectrum predicted by the hard pomeron model could
be observed. In a recent publication the exclusive production rate for
$ n $ jets has been calculated in this model [ 2 ]. Since it is not
infrared stable, a cutoff on the lower limit of the transverse jet
momentum
 had to be introduced. As a result, analytic calculations
became impossible. Numerical results presented in [ 2 ] seem to depend
crucially on the introduced infrared cutoffs both for real and virtual
gluons.

In this note we draw attention to the fact that the
$ n $-jet inclusive cross-sections in the hard pomeron model admit
analytic study. We calculate these cross-sections without any infrared
regularization and find them infrared stable, as expected. Of course,
in the infrared region the model cannot pretend to describe realistic
QCD phenomena. So, to apply the model to the latter, one has to exclude
this region in some way, which is important for integrated quantities,
such as rapidity distributions or multiplicity moments. To this purpose
introduction of some sort of cutoff is unavoidable. We however find
that the model itself provides for a certain infrared regularization
(although its origin is not quite conclusive, as will be discussed in
the following), so that it leads to finite rapidity distributions and
multiplicity moments.

Commenting on the results, we stress a characteristic
dependence of the
$ n $-jet production rate on the jet transverse momenta
$ k^{2}_{i} $. Since the
model is essentially logarithmic, it  predicts that all
$ \zeta_{i}=\ln k^{2}_{i} $ have the order of  at least
$ \sqrt{Y} $ where
$ Y=\ln s $ is the overall rapidity interval. 
It also predicts that all the differences
$ \zeta_{i}-\zeta_{j} $ have the same large order of magnitude.
 This means that in the model jet configurations are
favoured with transverse momenta of a widely different magnitude.
Due to dispersion in $\ln k^2$, the momenta of the observed jets result
strongly ordered, growing from the projectile or target regions towards
the center region. Also strong positive correlations are predicted both
in the forward-backward and forward-forward directions.
 This features could be
used to seek for an experimental evidence of hard pomerons.

The contents of this note is arranged as follows. In Section 2 we
deduce a general formula for the
$ n $-jet inclusive cross-section as a function of jet rapidities and
transverse momenta in the form suitable for further analytic study. In
Section 3  we study the single and double jet inclusive cross-sections
in some detail, having in mind their practical importance. Section 4 is
dedicated to a discussion of higher order inclusive cross-sections,
mostly in certain asymptotic regions of the phase space. Section 5
presents some conclusions.

%%%%%%%%%%%%%%%%%%%%%%%%%%%%%%%%%%%%%%%%%%%%%%%%%%%%%%%%%%%%%%%%%%%%%%%
\section{The
$ n $-jet inclusive cross-sections}
%%%%%%%%%%%%%%%%%%%%%%%%%%%%%%%%%%%%%%%%%%%%%%%%%%%%%%%%%%%%%%%%%%%%%%%
In the framework of the hard pomeron model jets are currently
identified with emitted real gluons. The
$ n $-jet inclusive cross-section is then described by the diagram
shown in Fig. 1, where also some notations for the transverse momenta
and their relative energetic variables are introduced. The blobs marked
$ G $ denote  pomeron forward Green functions which describe its
propagation between neighbouring jets and also between the jets with a
maximal or minimal rapidities and the projectile or target,
respectively. The latter are characterized by their respective colour
densities
$ \rho_{1} $ and
$ \rho_{2} $.

Emission of a real gluon (jet) of transverse momentum
$ k $ is described by a vertex
\beq
\frac{12\alpha_{s}}{k^{2}}l^{2}{l'}^{2}(2\pi)^{2}\delta^{2}(l-l'-k)
\eeq
where
$ l $ and
$ l' $ are the reggeized gluon momenta before and after emission;
$ \alpha_{s} $ is the (fixed) strong coupling constant. Passing to the
coordinate space, this vertex transforms into
\beq
\frac{12\alpha_{s}}{k^{2}}\stackrel{\leftarrow}{\Delta}
 e^{ikr}\stackrel{\rightarrow}{\Delta}
\eeq
where the Laplacian operators act in the directions indicated by the
arrows. Combining these vertices with  the pomeron Green functions
and  densities
$ \rho$  we obtain for the
$ n $-jet inclusive cross-section
\[
I_{n}(y_{i},k_{i})\equiv\prod_{i=1}^{n}\frac{d^{3}}{dy_{i}d^{2}k_{i}}\sigma=
\frac{128\pi^2}{9}\alpha_{s}^{2}
\prod_{i=1}^{n}\frac{12\alpha_{s}}{4\pi^2 k_{i}^{2}}
\int d^{2}rd^{2}r'\rho_{1}(r)\rho_{2}(r')
\prod_{i=1}^{n}d^{2}r_{i}\]\beq
G_{s_{01}}(r,r_{1})\stackrel{\leftarrow}{\Delta_{1}}e^{ik_{1}r_{1}}
\stackrel{\rightarrow}{\Delta_{1}}G_{s_{12}}(r_{1},r_{2})\stackrel
{\leftarrow}{\Delta_{2}}...\stackrel{\rightarrow}{\Delta_{n}}
G_{s_{n,n+1}}(r_{n},r')
\eeq
Here
$ s_{i,i+1} $ denotes the energetic variable for the pomeron between
 the
$ i $-th and
$ i+1 $-th jets. In terms of  jet rapidities
$ y_{i} $ and transverse momenta
$ k_{i} $
\[
 s_{i,i+1}=|k_{i}||k_{i+1}e^{y_{i}-y_{i+1}},\ \ i\neq 0,n,
 \]\beq
 s_{01}=\sqrt{s}|k_{1}|e^{-y_{1}},\ \
 s_{n,n+1}=\sqrt{s}|k_{n}|e^{y_{n}}
 \eeq
 Evidently they satisfy
 \beq
 \prod_{i=0}^{n}s_{i,i+1}=s\prod_{i=1}^{n}k_{i}^{2}
 \eeq
 Eq. (3) is valid provided all energetic variables between jets are
 large and all momenta squared are much smaller: $ k_{i}^{2},k_{i+1}^{2}
 <<s_{i,i+1}\rightarrow\infty
$ . Then we approximately have
\beq
 \ln s_{i,i+1}=y_{i}-y_{i+1}\equiv y_{i,i+1}
 \eeq
 for all
$ i=0,..n $, with
$ y_{0}=-y_{n+1}=\ln\sqrt{s}=(1/2)Y $. Note however, that
 exponentiating (6) reproduces only the dominant factor
$ s$  on the right-hand side of (5).

 The BFKL pomeron forward Green function
$ G_{s}(r,r') $ is well-known [ 3 ]. In the limit
$ s\rightarrow\infty $ we can retain only its dominant, isotropic, part
 \beq
 G_{s}(r,r')=\frac{rr'}{32\pi^{2}}
\int\frac{d\nu s^{\omega(\nu)}}{(\nu^{2}+1/4)^{2}}
 (\frac{r}{r'})^{-2i\nu}
 \eeq
 Alternatively we may restrict ourselves to studying only the isotropic
 part of the
$ n $-jet cross-section obtained after the integration  over all
 azimuthal angles of
$ k_{i} $, which results in taking only the isotropic part (7) of all
Green functions. Applying the Laplacian operators we find
 \beq
 \Delta G_{s}(r,r')=
 \frac{1}{8\pi^{2}}\int\frac{d\nu s^{\omega(\nu)}}{(1/2+i\nu)^{2}}
 (\frac{r}{r'})^{-1-2i\nu}
\eeq
 and a similar formula for the Laplacian operator applied to
$ r' $. As a result
 \beq
 \stackrel{\rightarrow}{\Delta}G_{s}(r,r')\stackrel{\leftarrow}{\Delta'}=
 \frac{1}{2\pi^{2}rr'}\int d\nu s^{\omega(\nu)}
 (\frac{r}{r'})^{-2i\nu}
\eeq
 Putting this into (3) we additionally introduce integrations over
 ${n+1}$  variables
$ \nu_{i,i+1},\ i=0,...n $.

 At each emission vertex we then find an integral over the transverse
 distance
$ r $ between the gluons
 \beq
 \int d^{2}r r^{-2-2i\nu} e^{ikr}=\frac{\pi i}{\nu+i0}(k/2)^{2i\nu}
 \frac{\Gamma (1-i\nu)}{\Gamma(1+i\nu)}
 \eeq
 where for the
$ i $-th jet
$ \nu=\nu_{i-1,i}-\nu_{i,i+1} $. The
$ n $-jet cross-section takes the form
 \[
I_{n}(y_{i},k_{i}^{2})\equiv
\prod_{i=1}^{n}\frac{d^{3}}{dy_{i}dk_{i}^{2}}\sigma=
\frac{4}{9}\alpha_{s}^{2}
\prod_{j=1}^{n}\frac{12i\alpha_{s}}{8\pi^{2} k_{j}^{2}}
\]\[\int d^{2}rd^{2}r'\rho_{1}(r)\rho_{2}(r')rr'
\prod_{i=0}^{n}d\nu_{i,i+1}s_{i,i+1}^{\nu_{i,i+1}}
 \frac{r^{-2i\nu_{01}}}{(1/2+i\nu_{01})^{2}}
  \frac{r^{+2i\nu_{n,n+1}}}{(1/2-i\nu_{n,n+1})^{2}}\]\beq
\prod_{i=1}^{n}\frac{(k_{i}^{2}/4)^{i(\nu_{i-1,i}-\nu_{i,i+1})}}
 {\nu_{i-1,i}-\nu_{i,i+1}+i0}\frac{\Gamma (1-i(\nu_{i-1,i}-\nu_{i,i+1}))}
 {\Gamma (1+i(\nu_{i-1,i}-\nu_{i,i+1}))}
\eeq

 Now we recall that the expression (11) is valid only for large enough
 relative energies
$ s_{i,i+1} $. The exponent
$ \omega(\nu) $ has a maximum at
$ \nu=0 $:
 \beq
 \omega (\nu)=\Delta-a\nu^{2}+{\cal O}(\nu^{4})
 \eeq
 where
 \beq
 \Delta=(12\alpha_{s}/\pi)\ln 2
 \eeq
 is the pomeron intercept and
 \beq
 a=(42\alpha_{s}/\pi)\zeta(3)
 \eeq
 Evidently at large
$ s_{i,i+1} $ small values of all
$ \nu_{i,i+1} $ give the dominant contribution. Consequently we can neglect
$ \nu $'s in all places where it is not accompanied by some large
 factors. The two integrations over
$ r $ and
$ r' $ then give average transverse dimensions of the projectile and
 target
 \beq
 r_{01(02)}=\int d^{2}r r \rho_{1(2)}(r)
 \eeq
 Representing
 \[
 \frac{k^{2i\nu}}{\nu+i0}=-i\int_{\ln k^{2}}^{\infty}\frac{d\alpha}{2\pi}
 e^{i\alpha\nu}\]
 we can integrate over all
$ \nu$'s introducing
$ n $ integrations over
$ \alpha_{i}, \ i=1,...n $ instead. Then we obtain for the
$ n $-jet cross-section
  \beq
I_{n}(y_{i},k_{i}^{2})=
\frac{64}{9}\alpha_{s}^{2}
R_{1}R_{2}\prod_{i=1}^{n}\frac{3\alpha_{s}}{4\pi^{3}
k_{i}^{2}}(s\prod_{i=1}^{n}k_{i}^{2})^{\Delta}
(\sqrt{\pi/a})^{n+1}\prod_{i=0}^{n}y_{i,i+1}^{-1/2}\,J_{n}
  \eeq
  where
  \beq J_{n}=
\prod_{i=1}^{n}\int_{\zeta_{i}}^{\infty}d\alpha_{i}
\exp
\left(-\sum_{j=0}^{n}\frac{(\alpha_{j}-\alpha_{j+1})^{2}}
{4ay_{j,j+1}}\right)
 \eeq
 and we have used the notation
  \beq
  \zeta_{i}=\ln k_{i}^{2}
  \eeq
  and have assumed
$ \alpha_{0}=\alpha_{n+1}=0 $

  Expression (16)-(17) is still far from being explicit, since integrations
 over
$ \alpha $'s cannot be easily done, except for the simplest case
$ n=1 $. However the integrand is now a positive function well behaved
 at all values of
$\alpha $'s. Therefore Eqs. (16)-(17) can be conveniently used both for
 numerical calculations and analytic study in certain limiting
 regions of jet parameters. Before we pass to the latter study
 we want to point out some evident general features of the
$ n $-jet cross-section (16)-(17).

It is clear that its overall behaviour in
$ s $ is independent of
$ n $ and is the same as for the total cross-section, namely,
$ \sim s^{\Delta}/\sqrt{\ln s} $. Indeed the characteristic values
of all $ \alpha $'s are of the order
$ \sqrt{y_{i,i+1}}\sim\sqrt{\ln s} $. Therefore integrations over
$ \alpha $'s will give a factor
$ \sim (\ln s)^{n/2} $, which will cancel a similar factor in (16)
 leaving one
$ \sqrt{\ln s} $ in the denominator. Passing to the behaviour in
$ k_{i}^{2} $, that is, in
$ \zeta_{i} $, we observe that at large
$ \zeta_{i} $ the exponentials in the integrand provide for a damping
 factor at $\zeta_{i}>\sqrt{Y}$. As a result all
$ \zeta $'s and also their differences
$ |\zeta_{i}-\zeta{j}| $ have large average values of the order
$ \sqrt{Y} $. This leads to a distribution in
$ k_{i}^{2} $ of the form discussed in the Introduction:
 neighbouring jets prefer to have their transverse momenta widely
 different in magnitude.

 Of special interest is the infrared behaviour of (16)-(17), when all
$ \zeta_{i}\rightarrow -\infty $. The integrals over
$ \alpha $'s converge to a finite value in this limit. Therefore the
 behaviour of the cross-section in each
$ k_{i}^{2} $ will be determined by the factor
$ k_{i}^{2\Delta}/k_{i}^{2} $. Taken at its face value, it leads to
 the cross-section (16) integrable at small $k_{i}^{2}$  and thus
 infrared stable. However we have to acknowledge that retaining the
 factor
$ k_{i}^{2\Delta} $ in our formulas is in fact beyond the accuracy
 adopted in the lowest order hard pomeron model, in which terms
 containing powers of
$ \alpha_{s}\ln k^{2} $ have been systematically neglected.
 Actually the region of small
$ k_{i}^{2} $ is not only beyond the applicability of the hard pomeron
 model to the realistic QCD but also beyond the validity of the model
 itself, since, as follows from (4), it corresponds to relative energies
$ s_{i,i+1} $ becoming small, when the lowest order approximation
 fails. For this reason the small
$ k_{i}^{2} $ region has to be excluded in any case, whether one wants
 to apply the model to the realistic QCD or to study the model as it
stands. The factors
$ k^{\Delta} $ which have appeared in (16) serve precisely to this
purpose: they effectively cut off the small
$ k_{i}^{2} $ region. Of course, one may alternatively use harder
cutoffs, as currently applied in the hard pomeron calculations, when
the integration over
$ k^{2} $ is simply cut off at some minimal
$ k_{min}^{2} $. Since the difference, if any, reflects only the
unknown influence of the infrared region, in the following
we shall retain the factors
$ k^{2\Delta} $ in (16) as a natural infrared cutoff.

Finally we note that $J_n$, Eq.(17), can be rewritten as
  \beq J_{n}=
\prod_{i=1}^{n}c_{i}\int_{\zeta_{i}/c_{i}}^{\infty}d\alpha_{i}
\exp
(-\sum_{j=1}^{n}\alpha_{j}^{2}-2\sum_{j=1}^{n-1}
b_{j}\alpha_{j}\alpha_{j+1})
 \eeq
where
\[ c_{i}=\sqrt{\frac{4ay_{i-1,i}y_{i,i+1}}{y_{i-1,i+1}}},\ \ 
b_{i}=\sqrt{\frac{y_{i-1,i}y_{i+1,i+2}}{y_{i-1,i+1}y_{i,i+2}}} \]
and we have denoted $y_{i,i+2}=y_{i,i+1}+y_{i+1,i+2}$. From
this expression it is clearly seen that correlations are predicted 
between  jets neighbouring in rapidity (i.e. $i$ and $i+1$).
Their strength is determined by the value of $b_j$'s, which depends
on the rapidity distances between the correlating jets and their 
neighbours. Evidently the correlations are maximal if the jets are emitted 
at the center of the rapidity interval between their neighbours and vanish
if they are emitted at its borders. One also sees that the correlations 
are positive if the two jets have large momenta ($\zeta_i,\zeta_{i+1}>0$),
whereas they may be both positive and negative if one or both jets have
very small momenta ($\zeta_i$ or/and $\zeta_{i+1}<0$). Study of the
2-jet correlations shows that in the latter case the correlations are
negligible.
%%%%%%%%%%%%%%%%%%%%%%%%%%%%%%%%%%%%%%%%%%%%%%%%%%%%%%%%%%%%%%%%%%%%%%%%%%%
\section{Single and double inclusive cross-sections}

Although single jet inclusive cross-sections have already been studied
rather extensively since long ago, we found it convenient to start with
them to clearly see the effect of the natural cutoff provided by the
$k^{2\Delta}$ factor in (16). With it the production rate can be
written explicitly. Earlier studies mostly used harder cutoffs, like
just excluding the region below some chosen minimal value $k_{min}^{2}$.

The single jet inclusive cross section, obtained from (16) for $n=1$
can be written in terms of the error function
\beq
I_1(y_1,k^2_1)=\frac{3\alpha_s}{4\pi^2}\sigma
k_1^{2(\Delta-1)}\left(1-\Phi(x_1)\right)
\eeq
where
\beq
x_1=\zeta_1\sqrt{\frac{Y}{4ay_{01}y_{12}}}
\eeq
and $\sigma$ is the total cross section
corresponding to a pomeron exchange, which is a common factor to all
inclusive cross-sections:
\beq
\sigma=\frac{64}{9}\alpha_s^2 R_1R_2s^{\Delta}\sqrt{\frac{\pi}{aY}}
\eeq
For finite $k_1^2$ and thus finite $\zeta_1$, the argument $x_1$ is small
and the behavior of $I_1$ in $k^2$ is totally determined by the factor
$1/k_1^2$. At very small $k_1^2 $ ($\zeta_1\rightarrow -\infty$)
the cross-section is cut by the factor $k_1^{2\Delta}$ at values
$\zeta_1\sim -1/\Delta$. At very large $k_1^2$ a cutoff is provided 
by the error function at $\zeta_1\sim\sqrt{4ay_{01}y_{12}/Y}$

Integrating (20) over $k_1^2$ one obtains the distribution in rapidity
\beq
I_1(y_1)=\frac{3\alpha_s}{2\pi^2\Delta}\sigma\exp (a\Delta^2 y_{01}
y_{12}/Y)
\eeq
Since $y_{01}y_{12}=Y^2(\frac{1}{4}-y_1^2)$ this distribution is
evidently peaked at the center rapidity. 

One should note that the
exponent in (23) has the order $\alpha_{s}^3 Y$ and thus, strictly speaking
should be neglected in the lowest order approximation, in which only terms
of the order $\alpha_s Y$ has been taken into account. Keeping the 
exponent has the same meaning as
retaining the factors $k^{2\Delta}$ in the initial formulas and should be
considered as a natural cutoff provided by the kinematical constraints
in the theory.

To find an average value of the jet momentum at a given rapidity
one has to integrate (20)
with the weight $\zeta_1$ and divide by (23). One then obtains
\beq\langle \zeta_1\rangle=(1/2)a\Delta (Y-4y_1^2/Y)
\eeq
As expected from the BFKL diffusion, the average momentum
grows from the edges of the rapidity interval towards the center.
The characteristic order of $\zeta_1=\ln k_1^2$ results to be 
$\alpha_s^2 Y$, and not $\sqrt{Y}$ as deduced from the BFKL wave
function properties. This also is a consequence of retaining the
$k^{2\Delta}$ factor, which evidently somewhat enhances the observed
jet momentum.

Integrating (23) over all rapidity range one obtains the total multiplicity as
\beq
\langle n\rangle= \frac{3\alpha_s}{2\pi^2\Delta^2}\sqrt{\frac{\pi Y}{a}}
\exp(a\Delta^2 Y/4) \Phi\left(\frac{1}{2}\Delta\sqrt{aY}\right)
\eeq
For $\Delta\sqrt{aY}<<1$ the average multiplicity grows linearly with $Y$,
as expected. However if one takes $Y\rightarrow\infty$ keeping $\Delta$
fixed, $\langle n\rangle$ rises as a power of energy $\sim s^{a\Delta^2/4}$.

Now we pass to the double inclusive cross-section given by (16) for $n=2$:
\beq 
I_2(y_1,k_1^2;y_2,k_2^2)=
\frac{9\alpha_s^2}{16\pi^5a}\sigma
(k_1k_2)^{2\Delta-1}
\sqrt{\frac{Y}{y_{01}y_{12}y_{23}}}J_{2}\eeq
where
\beq
J_{2}=\int_{\zeta_1}^{\infty}d\alpha_1
\int_{\zeta_2}^{\infty}d\alpha_2\exp \left(-\frac{\alpha_1^2}{4ay_{01}}-
\frac{(\alpha_1-\alpha_2)^2}{4ay_{12}}-\frac{\alpha_2^2}{4ay_{23}}\right)
\eeq
As well-known, this cross-section was studied in [4] under different kinematical 
conditions
(small $y_{01}$ and $y_{23}$ and large $y_{12}$). Only the dependence 
on $y_{12}$
 was discussed as a possible experimental signature of the BFKL pomeron 
 between the two jets.

As seen from (19) real parameters of the integral (27) are
$\zeta_1/\sqrt{4aY}$ and $\zeta_2/\sqrt{4aY}$. So for finite
jet momenta one can put $\zeta_1=\zeta_2=0$ and the integral, which turns
out independent of jet momenta, can be easily calculated;
\[J_{2}=\pi a\sqrt{y_{01}y_{12}y_{23}/Y}
\left(1+\frac{2}{\pi}\arctan
\sqrt{\frac{y_{01}y_{23}}{y_{12}Y}}\right)\]
so that the inclusive cross-section becomes
\beq 
I_2(y_1,k_1^2;y_2,k_2^2)=
\frac{9\alpha_s^2}{16\pi^4}\sigma
(k_1^2k_2^2)^{\Delta-1}
\left(1+\frac{2}{\pi}\arctan
\sqrt{\frac{y_{01}y_{23}}{y_{12}Y}}\right)
\eeq

For general $\zeta_1$ and $\zeta_2$ 
the integral (27), which is an evident generalization of the error 
function to two dimensions, cannot be done analytically 
and requires
numerical calculation.  
We have calculated the ratio
\beq
R(y_1,k_1^2;y_2,k_2^2)=\frac{\sigma I_2(y_1,k_1^2;y_2,k_2^2)}
{I_1(y_1,k_1^2)I_1(y_2,k_2^2)} 
\eeq
to study correlations between the emitted jets. Absence 
of correlations corresponds to $R=1$. In Fig. 2 we show results of 
the numerical calculation of $R$ for a symmetric pair of jets at $y_1=-
y_2=Y/6$, which divide the rapidity interval into three equal parts.
$R$ is presented as a function of $z_2=\zeta_2/\sqrt{4aY}$ in the interval
$0\leq z_2\leq 2$ for fixed positive values of $z_1=\zeta_1/\sqrt{4aY}<1$. 
 For negative $\zeta_1$ or/and $\zeta_2$ we find that $R=1$ and there are
 no correlations. One observes from Fig. 2 that there are 
 significant positive correlations between the two jets
 in the whole region 
 of positive 
 $\zeta_1$ and $\zeta_2$  corresponding to momenta which are not 
 small relative to the intrinsic scale of the BFKL model (which should
 be supplied by higher orders of the theory). They should be seen 
 experimentally as positive forward-backward correlations in the c.m. system.

 For very large jet momenta, when $\zeta_1$ or/and $\zeta_2$ become of the
 order $\sqrt{4aY}$, the correlations become very strong and grow very fast 
 with $\zeta$'s, reaching values of the order $4.\,10^5$ at $\zeta_1=\zeta_2
 =2$. However one should have in mind that at such high momenta 
 the production rate gets exceedingly small, beyond any possibility of
 experimental detection, which can be deduced from (20), so that such high
 values of $R$ bear, in fact, no physical importance.

 We have also calculated the ratio $R$ for two jets emitted both in the
 forward hemisphere, with $y_1=Y/3$ and $y_2=Y/6$. The results are 
 quite similar qualitatively, although the numerical values result a bit
 different (however of the same order of magnitude). Thus one should also
 expect forward-forward correlations of the same order as 
 the forward-backward ones.

 Integrating (26) over the two momenta $k_1^2$ and $k_2^2$ one obtains the double
 distribution in rapidity. These integrations can easily be done explicitly
 using an evident formula
\beq
\int_{-\infty}^{+\infty} d\zeta e^{\Delta\zeta}\int_{\zeta}^{+\infty}
d\alpha f(\alpha)=(1/\Delta)\int_{-\infty}^{+\infty}d\zeta
e^{\Delta\zeta}f(\zeta)
\eeq
valid for any
$ f(\alpha) $ integrable at
$ \alpha=\pm\infty $. We obtain 
\beq
I_2(y_1,y_2)=\left(\frac{3\alpha_s}{2\pi^2\Delta}\right)^2 \sigma
\exp\left(a\Delta^2(y_{01}+y_{23}-(y_{01}-y_{23})^2/Y)\right)
\eeq

The exponential factor again comes from the kinematical
factor $k^{2\Delta}$ and has a higher order than normally taken into
account in the BFKL theory. If we neglect it, we obtain a completely
flat distribution in the two rapidities.  
The exponential factor evidently has its maximum when the two jets are both
emitted at central rapidity $y_1=y_2=0$, when it grows like 
$\exp(a\Delta^2 Y)$.
Comparing to (23) we  observe that  it also introduces positive
correlations between the two jets in the rapidity space. In fact, the ratio
analogous to (29) results
\beq
R(y_1,y_2)=\frac{\sigma I_2(y_1,y_2)}{I_1(y_1)I_1(y_2)}=
\exp (a\Delta^2 y_{01}y_{23}/Y)
\eeq
and evidently implies positive correlations, growing as the two jets move
to the central rapidity region.

Integrating (31) over all
the rapidity region we obtain  the second moment of the distribution 
in the number of jets. Neglecting the exponential factor we get
\beq
\langle n(n-1)\rangle= 
\left(\frac{3\alpha_s}{2\pi^2\Delta}\right)^2 Y^2
=\langle n\rangle^2
\eeq
With the exponential factor taken into account, the second moment can
be found only by numerical integration. 
%%%%%%%%%%%%%%%%%%%%%%%%%%%%%%%%%%%%%%%%%%%%%%%%%%%%%%%%%%%%%%%
\section{Multijet inclusive cross-sections}
The features found for the production of two jets repeat 
themselves in the multijet
production. However for $n>2$ analytic results
cannot be obtained for most of the observables. So one can only study
cases corresponding to some limiting parts of the phase space.

\subsection{Distributions in rapidities and momenta}

As mentioned the real parameters of the cross-section are ratios
$ \zeta_{i}/\sqrt{4aY} $. 
Depending on their values we may
distinguish between three asymptotic regions, in which
the cross-section (16) can be estimated by standard asymptotic methods.

The first region is the infrared one, when
$ \zeta_{i}\rightarrow -\infty $ faster than
$ \sqrt{Y}\rightarrow\infty $. In this region the cross-section
is given by (16) with the integrals
$ J_{n} $ independent of
$ k_{i}^{2} $ and given by
  \beq J_{n}=
\prod_{i=1}^{n}\int_{-\infty}^{\infty}d\alpha_{i}
\exp
\left(-\sum_{j=0}^{n}\frac
{(\alpha_{j}-\alpha_{j+1})^{2}}{4ay_{j,j+1}}\right)
 \eeq
To do this integral we pass to variables
\beq
\alpha_{i,i+1}=\alpha_{i}-\alpha_{i+1}, \ \ i=0,...n
\eeq
subject to condition
$ \sum_{i=0}^{n}\alpha_{i,i+1}=0 $ and present
\beq
J_{n}=\int_{-\infty}^{\infty}\frac{d\lambda}{2\pi}
\prod_{i=0}^{n}\int_{-\infty}^{\infty}d\alpha_{i,i+1}
\exp
\sum_{j=0}^{n}\left(i\lambda
-\frac{\alpha_{j,j+1}^{2}}{4ay_{j,j+1}}\right)
 \eeq
All integrals are trivial and we get
\beq
J_{n}=\frac{1}{2\pi}\sqrt{\frac{\pi}{aY}}\prod_{i=0}^{n}(4\pi
ay_{i,i+1})^{1/2}
\eeq
Thus in the infrared region we have the cross-section
  \beq
I_{n}(y_{i},k_{i}^{2})=
(\frac{3\alpha_{s}}{2\pi^{2}})^n \sigma
(\prod_{i=1}^{n}k_{i}^{2})^{\Delta-1}
  \eeq
As one observes, the infrared cross-section is independent of
individual jet rapidities: the distribution in all
$ y_{i} $ is completely flat. The dependence on
$ k_{i}^{2} $ apart from the canonical factor
$ 1/k_{i}^{2} $ exhibits an extra factor discussed above
$ k_{i}^{2\Delta} $, which serves as an effective infrared cutoff.

The second region is that of low
$ k_{i}^{2} $, when
$ \zeta_{i} $ are finite and thus all ratios
$ \zeta/\sqrt{Y} $ go to zero. In this region the inclusive
cross-section is  given by (16) with the integral
$ J_{n} $ again independent of
$ k_{i}^{2} $ :
  \beq J_{n}=
\prod_{i=1}^{n}\int_{0}^{\infty}d\alpha_{i}
\exp
\left(-\sum_{j=0}^{n}\frac{(\alpha_{j}-\alpha_{j+1})^{2}}
{4ay_{j,j+1}}\right)
 \eeq
The dependence on
$ k_{i}^{2} $ is thus the same as in the infrared region. However there
appears now a nontrivial dependence on rapidities. 

The integral (39) can be easily done only for 
$ n=1,2 $ (see Sec. 3). 
Calculation of
$ J_{n} $ for larger
$ n$  can be done by numerical methods. Analytic estimate of
$ J_{n} $ is possible within the saddle point method, presumably valid
for large
$ n$. 
Denote
\beq
P=\sum_{i=0}^{n}\frac{(\alpha_{i}-\alpha_{i+1})^{2}}{4ay_{i,i+1}}\equiv
\sum_{i=0}^{n}b_{i,i+1}(\alpha_{i}-\alpha_{i+1})^{2}
\eeq
 We have
\beq
\sum_{i=k+1}^{k+m}\partial P/\partial
\alpha_{i}=2b_{k,k+1}(\alpha_{k+1}-\alpha_{k})+
2b_{k+m,k+m+1}(\alpha_{k+m}-\alpha_{k+m+1})\eeq
If we require that
\beq\partial P/\partial \alpha_{i}=0,\eeq
for all $ i=1,...n$
then, recalling that
$ \alpha_{0}=\alpha_{n+1}=0 $, we shall obtain an homogeneous system of
linear equations in
$ \alpha_{i}, i=1,...n $ with a nonsingular matrix, so that the
solution will be
that all
$ \alpha $ are zero, i.e. at the initial integration point.
Thus there are no saddle points inside the overall integration volume
in all
$ \alpha $'s. We can, however, relax condition (42), by fixing one of the
$ \alpha $'s, say
$ \alpha_{n} $, thus seeking a saddle point on the surface, 
inside the integration volume in all other
$ \alpha_{i},\ \ i=1,...n-1 $. Putting 
$ k+m=n-1 $ in (41), we get
\beq
\alpha_{k}-\alpha_{k+1}=
-\frac{b_{n-1,n}}{b_{k,k+1}}(\alpha_{n-1}-\alpha_{n})
\eeq
Summing these equations for
$ k=0,..j-1 $ we obtain
\beq
-\alpha_{j}=b_{n-1,n}(\alpha_{n-1}-\alpha_{n})
\sum_{k=0}^{j-1}1/b_{k,k+1}\equiv
\frac{b_{n-1,n}}{b_{0j}}(\alpha_{n-1}-\alpha_{n})
\eeq
Here and in the following we use the notation
\beq
1/b_{j,j+m}=\sum_{k=j}^{j+m-1}1/b_{k,k+1}=4a(y_j-y_{j+m})
\eeq
Putting
$ j=n $ in (44), we can express
the difference   $\alpha_{n-1}-\alpha_{n} $  in
terms of the fixed variable
$ \alpha_{n} $. Combining this with (44) we finally find
\beq
\alpha_{j}=\frac{b_{0,n}}{b_{0,j}}\alpha_{n}
\eeq
 Eqs. (46) determine the position of the saddle
point in $ \alpha_{i},\ i=1,...n-1 $ in terms of the fixed $ \alpha_{n}
$. At the saddle point we find \beq
P=\sum_{i=0}^{n-1}b_{i,i+1}\left(\frac{b_{n-1,n}}{b_{i,i+1}}
(\alpha_{n-1}-\alpha_{n})\right)^{2}+b_{n,n+1}\alpha_{n}^{2}
=\alpha_{n}^{2}(b_{0n}+b_{n,n+1}).
\eeq
 Integration over the saddle point gives a factor
\[ \sqrt{\frac{\pi^{n-1}}{\mbox{det}\,M}}\]
where
$ M$ is an
$ (n-1)\times(n-1)$ matrix of second derivatives.
Simple calculations give
\beq
\mbox{det}\,M=
(1/b_{0,n})\prod_{i=0,n-1}b_{i,i+1}
\eeq
The  integration over
$ \alpha_{n} $ adds a factor
$\frac{1}{2}\sqrt{\pi/(b_{0n}+b_{n,n+1})}$.
and we get  the contribution from the surface $\alpha_n=0$
\beq
J_{n}=\frac{1}{4\pi}\sqrt{\frac{\pi}{aY}}
\prod_{i=0}^{n}(4\pi ay_{i,i+1})^{1/2}
\eeq
which is one half of (37). Taking into account all other surface 
contributions will multiply (49) by $n$.
 As a result, we obtain for the inclusive
cross-section an expression which is identical to (38) except for
an extra factor
$ n/2 $. The distribution in rapidities obtained by this asymptotic
estimate is again completely flat.

We finally pass to the third asymptotic region, that of high
$ k_{i}^{2} $, which corresponds to all
$ \zeta_{i}/\sqrt{y}\rightarrow\infty $. In this region the behaviour
of the integral (17) is governed by the damping exponentials in the
integrand.
 To find the asymptotics we may again recur to the saddle point method.
However in this region the integration volume depends on the values of
$\zeta_{i} $ and we have to check whether the found saddle point
lies inside this volume. As a result, the set of
$ \alpha $'s for which the saddle point lies inside the integration
volume will depend on the values of
$ \zeta_{i} $ and will generally be only a part of all
$ \alpha $'s. The rest of
$ \alpha $'s will be restricted to their initial values
$ \zeta $'s, since in absence of a saddle point the integrand
falls with each
$ \alpha $ exponentially. The asymptotics will thus depend in a
rather complicated way on the manner in which different
$ \zeta $'s get large and also on the jet rapidities. Nevertheless
it is not difficult to find this asymptotics for given
$ y_{i} $ and large
$ \zeta_{i} $.

Let us consider a situation when a particular set of $\alpha$'s
take on their initial values in the asymptotics, the rest of them
integrated over their saddle points. Take a pair of jets, the
$ j$ th and
$ (j+m) $th, for which
$ \alpha $'s take on their initial values
 with all intermediate jets
from
$ (j+1) $th to
$ (j+m-1) $th having a saddle point inside the integration volume.
In other words Eqs. (42) are satisfied for
$ i=j+1,...j+m-1 $ with $\alpha_{j}=\zeta_{j}
$  and
$ \alpha_{j+m}=\zeta_{j+m} $. Repeating the procedure described above
we can solve these equations  for the saddle point
$ \alpha_{i}, i=j+1,...j+m-1 $ in terms of
$ \zeta_{j} $ and
$ \zeta_{j+m-1} $. We find
\beq
\alpha_{j+k}=\zeta_{j}+\frac{b_{j,j+m}}{b_{j,j+k}}(\zeta_{j+m}-\zeta_{j})
>\zeta_{j+k},\ \ k=1,...m-1
\eeq
The last inequality follows from the condition that the saddle point
should lie inside the integration volume. This inequality determines a
certain domain in the jet phase space
$ y_{i}, k^{2}_{i} $ where this particular type of asymptotic behaviour
will take place. The part of the exponential
$ P $ containing
$ \alpha $'s from the
$ (j+1) $th to
$( j+m-1) $th, at the saddle point becomes equal to
\[ b_{j,j+m}(\alpha_{j}-\alpha_{j+m})^{2}\]
that is, takes the same form as if there were no other jets between the
$ j $ th and $(j+m)
$  th  ones. Integration over all $\alpha_{i},\ i=j+1,... j+m-1
$  by the saddle point method thus gives a factor
\beq
\sqrt{\frac{\pi^{m-1}b_{j,j+m}}{\prod_{i=j}^{j+m-1}b_{i,i+1}}}\eeq
and leaves an integrand where these jets have disappeared. Doing
this with all sets of
$ \alpha $'s which have their saddle point inside the integration volume
we shall obtain more factors (51) and an integral over remaining
$ \alpha$'s in which all other jets have disappeared and which does
not contain saddle points inside the integration volume. Integrating
by parts one obtains for the latter integral an asymptotics
\beq \frac{1}{\prod \partial P/\partial\alpha_{i}}_{\alpha_{i}=\zeta_{i}}
\exp\left(-\sum_{i<k} b_{i,k}(\zeta_{i}-\zeta_{k})^{2}\right)
\eeq
Both the product and sum go over the jets remaining after exclusion of
the ones integrated over their saddle points. The described asymptotics
will evidently be valid in the domain where the inequality (50) is
fulfilled for these excluded jets.

To make this rather complicated prescription more understandable,
consider a simple case when rapidity distances between neighbouring
jets are all equal:
$  y_{i,i+1}=\Delta y $ independent of
$ i $. Then the left hand side of inequality (50) represents
 a direct line in the
$ i,\zeta $ plane connecting points
$ \zeta_{j} $ and
$ \zeta_{j+m} $. The inequality (50) means that the intermediate
$ \zeta_{i},\ i=j+1,...j+m-1 $ lie below this line. The prescription
for the asymptotics is  then to draw all lines connecting given
$ \zeta_{i}, \ i=0,1,..n,n+1 $, where by definition
$ \zeta_{0}=\zeta_{n+1}=0 $. If some
$ \zeta $'s happen to lie below any such line, one should throw
the corresponding jets out introducing factors (51) instead. In the end
only
$ \zeta $'s will remain which lie above any line connecting
 a pair of them, that is the remaining
$ \zeta_{i} $ will represent a convex line in the
$ i,\zeta $ plane. The behaviour in these
$ \zeta $'s  will be described by (52). Of course, if
if all $ \zeta_{i} $ form a convex line from the start, the asymptotics
will have the form (52) with a product and sum going over all
$ i=1,...n $.

For illustration consider a case
$ n=2 $. Condition that points
$ \zeta_{0}=0,\zeta_{1},\zeta_{2},\zeta_{3}=0 $ form a convex function
of
$ i $ is evidently
\beq
|\zeta_{1}-\zeta_{2}|<\zeta_{1},\ \zeta_{2}
\eeq
In this region  the asymptotics will be
\beq
I_{2}=\frac{3\sqrt{3}\alpha_{s}^{2}}{4\pi^{5}}aY
\sigma (k_{1}^2k_{2}^2)^{\Delta-1}
\frac{1}{\zeta_{1}\zeta_{2}-2(\zeta_{1}-\zeta_{2})^{2}}
\exp\left(-\frac{\zeta_{1}^{2}+\zeta_{2}^{2}+(\zeta_{1}-\zeta_{2})^{2}}
{(4/3)aY}\right)
\eeq
Now take the region
\beq
2\zeta_{1}<\zeta^{2}
\eeq
where
$ \zeta_{1} $ lies  below the line connecting
$ \zeta_{0}=0 $ and
$ \zeta_{2} $. According to our prescription  the first
jet now gives a factor (51) and the second gives (52). We thus obtain
\beq
I_{2}=\frac{27\sqrt{2}\alpha_{s}^{2}}{8\pi^{4}}
sqrt{\frac{aY}{\pi}}\sigma
(k_{1}^2k_{2}^2)^{\Delta-1}
\frac{1}{\zeta_{2}}\exp\left(-\frac{\zeta_{2}^{2}}{4aY}\right)
\eeq
In the symmetric region
$ 2\zeta_{2}<\zeta_{1} $ we shall evidently obtain (56) with the first
and second jets interchanged.

\subsection{Integrated distributions}

Integrating (16) over all
$ k_{i}^{2} $ we get rapidity distributions
$ I_{n}(y_{i}) $. As discussed, these integrations converge in the
infrared region due to the cutoff factors
$ k_{i}^{2\Delta} $. Using   (30)
we obtain from (16)
\beq
I_{n}(y_{i})=
(\frac{3\alpha_{s}}{4\pi^{2}\Delta})^{n}
(\sqrt{\frac{\pi}{a}})^{n}\sigma
\sqrt{Y} \prod_{i=0}^{n}y_{i,i+1}^{-1/2}\,K_{n}
\eeq
where
\beq
K_{n}=\int_{-\infty}^{+\infty}\prod_{i=1}^{n}d\zeta_{i}
\exp\left(\Delta\sum_{i=1}^{n}\zeta_{i}-
\sum_{i=0}^{n}b_{i,i+1}(\zeta_{i}-\zeta_{i+1})^{2}\right)
\eeq
and
$ \zeta_{0}=\zeta_{n+1}=0 $ and
$ b_{i,i+1} $ are defined by (40). The integral (58) can be easily
calculated using the same technique that was applied to obtain (37).
In terms of variables
$ \zeta_{i,i+1}=\zeta_{i}-\zeta_{i+1} $, the sum of which is zero, we
find
\beq
\sum_{i=1}^{n}\zeta_{i}=(1/2)\sum_{i=0}^{n}(n-2i)\zeta_{i,i+1}
\eeq
The integral
$ K_{n} $ becomes
\[
K_{n}=\int_{-\infty}^{+\infty}\frac{d\lambda}{2\pi}
\prod_{i=1}{n}d\zeta_{i}
\exp
\sum_{i=0}^{n}\left((i\lambda+
(n/2-i)\Delta)\zeta_{i,i+1}-
b_{i,i+1}(\zeta_{i}-\zeta_{i+1})^{2}\right)=\]\beq
\frac{1}{2\pi}\prod_{i=0}^{n}(4\pi ay_{i,i+1})^{1/2}
\exp\left(\frac{a\Delta^{2}}{4Y}
(n^{2}Y^{2}+8\rho Y-4(n+1)\sigma Y-4\sigma^{2})\right)
\eeq
where
\beq
\sigma=\sum_{k=1}^{n}y_{k},\ \ \rho=\sum_{k=1}^{n}ky_{k}
\eeq
Putting (60) into (57) we obtain finally
\beq
I_{n}(y_{i})=
(\frac{3\alpha_{s}}{2\pi^{2}\Delta})^{n}
\sigma
\exp\left(\frac{a\Delta^{2}}{4Y}
(n^{2}Y^{2}+8\rho Y-4(n+1)\sigma Y-4\sigma^{2})\right)
\eeq

As in the two-jet case, the exponential factor lies, in fact, outside the
precision, normally adopted in the lowest order BFKL model. If we neglect it
we evidently obtain a completely flat
distribution in all rapidities. 
The effect of the exponential factor can easily be seen. Let
$ \phi=n^{2}Y^{2}+8\rho Y-4(n+1)\sigma Y-4\sigma^{2}$.
Since the derivatives
$ \partial\phi/\partial y_{k}=Y(8k-4(n+1))-8\sigma,\ k=1,...n $ cannot
all be equal to zero simultaneously, the maximum  of
$ \phi $ occurs at limiting values of its arguments, that is, when
all
$ y_{i} $ coincide:
$ y_{1}=y_{2}=...=y_{n} $. At this point
$ \phi=n^{2}(Y^{2}-4y_{1}^{2}) $. The maximum is achieved at
$ y_{1}=0 $ and is equal to
$ n^{2}Y^{2} $. Thus the exponential factor favours emission of all jets
at central rapidity transforming a flat plateau into a convex one.

Integration of
$ I(y_{i}) $ over all
$ y_{i} $ gives multiplicity moments.
With a flat distribution we immediately get
\beq
\langle n(n-1)...(n-k+1)\rangle=
(\frac{3\alpha_{s}}{2\pi^{2}\Delta})^{k}
Y^{k}=\langle n\rangle^k
\eeq
This corresponds to a Poissonian distribution in $n$,
with the average $\langle n\rangle$ growing like $Y$,
 in confirmation of simple estimates.
 The exponential factor in (62) will, in all probability, 
 make the moments grow more rapidly, like powers of energy,
 as follows from the case $n=1$ studied in Sec. 3
%%%%%%%%%%%%%%%%%%%%%%%%%%%%%%%%%%%%%%%%%%%%%%%%%%%%%%%%%%%%%%%%%%%%%%%%%%%
\section{Conclusions}
We have studied inclusive production of many jets in the hard pomeron model,
with a special emphasis on the two-jet production. 
In confirmation of current expectations, we found that the
average jet momenta are strongly ordered, monotonously growing 
from the projectile or target regions
to the central regions. The most characteristic feature
of multiple jet production seems to be strong positive correlations
between jets neighbouring in rapidity, both in the forward-backward
and forward-forward directions.

In our treatment we have preserved kinematical factors $k^{2\Delta}$, 
which serve as a natural infrared cutoff and allow to obtain finite
analytic expressions for  distributions in rapidity and number of jets. 
As a consequence we also obtained an enhancement in the high $k^2$
region, so that the average values $\langle \ln k^2 \rangle$ result
growing like $\alpha_s^2 Y$ and not like $\sqrt {\alpha_s Y}$ as
concluded from the study of the pomeron wave function. To see the
relevance of this effect, one has to study the second order corrections 
in the hard pomeron model, which now seems possible, but lies beyond
the scope of the present study.

%%%%%%%%%%%%%%%%%%%%%%%%%%%%%%%%%%%%%%%%%%%%%%%%%%%%%%%%%%%%%%%%%%%%%%%%%%%%
\section {Acknowledgments.}
%%%%%%%%%%%%%%%%%%%%%%%%%%%%%%%%%%%%%%%%%%%%%%%%%%%%%%%%%%%%%%%%%%%%%%%%%%%%
M.A. Braun thanks the INFN and the University of Trieste
for their financial help during his stay
at Trieste.
This work was partially supported by the Italian Ministry of University and of
Scientific and Technological Research by means of the Fondi per la Ricerca
scientifica - Universit\`a di Trieste.
%%%%%%%%%%%%%%%%%%%%%%%%%%%%%%%%%%%%%%%%%%%%%%%%%%%%%%%%%%%%%%%%%%%%%%%%%%%%
\section{References}
1. ZEUS Collaboration. M.Derrick {\it et al.}, Z.Phys. {\bf C69} (1996) 607.\\
   H1 Collaboration. T.Ahmed {\it et al.}, Nucl. Phys.,{\bf B439}(1995) 471.\\
2. J.Kwiecinski, C.A.M.Lewis and A.D.Martin, Phys. Rev. D 54 (1996) 6664.\\
3. L.N.Lipatov, in: "Perturbative QCD", ed. by A.H.Mueller, World Scientific,
Singapore (1989).\\
4. A.H.Mueller and H.Navalet, Nucl. Phys. {\bf B282} (1987) 727.
%%%%%%%%%%%%%%%%%%%%%%%%%%%%%%%%%%%%%%%%%%%%%%%%%%%%%%%%%%%%%%%%%%%%%%%%%%%%
\section{Figure captions}
Fig. 1. The diagram which describes the 
$ n $-jet inclusive cross-section.\\

\noi Fig. 2. The ratio $R$ (Eq. (29)) for two jets with $y_1=-y_2=Y/3$ as
a function of $z_2=\ln k^2_1/\sqrt{4aY}$ for different values
of $z_1=\ln k_1^2/\sqrt{4aY}$.\\
%%%%%%%%%%%%%%%%%%%%%%%%%%%%%%%%%%%%%%%%%%%%%%%%%%%%%%%%%%%%%%%%%%%%%%%%%%5

\end{document}